\documentclass[twocolumn,secnumarabic,amssymb, nobibnotes, aps, pre, showpacs]{revtex4-1}

\usepackage[dvipdfmx]{graphicx}
\usepackage{bm}
\usepackage{color}

\begin{document}

\title{Hawkes process model with a time-dependent background rate and its application to high-frequency financial data}

\author{Takahiro Omi}
\email{omi@sat.t.u-tokyo.ac.jp}
\affiliation{Institute of Industrial Science, The University of Tokyo, 4-6-1 Komaba, Meguro-ku, Tokyo 153-8505, Japan.}
\author{Yoshito Hirata}
\affiliation{Institute of Industrial Science, The University of Tokyo, 4-6-1 Komaba, Meguro-ku, Tokyo 153-8505, Japan.}
\author{Kazuyuki Aihara}
\affiliation{Institute of Industrial Science, The University of Tokyo, 4-6-1 Komaba, Meguro-ku, Tokyo 153-8505, Japan.}
\date{}

\begin{abstract}
A Hawkes process model with a time-varying background rate is developed for analyzing the high-frequency financial data.
In our model, the logarithm of the background rate is modeled by a linear model with a relatively large number of variable-width basis functions, and the parameters are estimated by a Bayesian method.
Our model can capture not only the slow time-variation, such as in the intraday seasonality, but also the rapid one, which follows a macroeconomic news announcement.
By analyzing the tick data of the Nikkei 225 mini, we find that (i) our model is better fitted to the data than the Hawkes models with a constant background rate or a slowly varying background rate, which have been commonly used in the field of quantitative finance; (ii) the improvement in the goodness-of-fit to the data by our model is significant especially for sessions where considerable fluctuation of the background rate is present; and (iii) our model is statistically consistent with the data.
The branching ratio, which quantifies the level of the endogeneity of markets, estimated by our model is 0.41, suggesting the relative importance of exogenous factors in the market dynamics.
We also demonstrate that it is critically important to appropriately model the time-dependent background rate for the branching ratio estimation.
\end{abstract}

\pacs{89.65.Gh, 05.45.Tp, 05.40.-a, 89.75.-k}

\maketitle

\section{INTRODUCTION}
In a variety of complex systems, the activity is driven by endogenous (internal) and exogenous (external) forces, exhibiting complex dynamics.
Such dynamics can be observed from natural science to social science, and the examples include earthquakes \cite{omori1894, iasacks1968, gardner1974, zhuang2002, marsan2008}, neuronal firing \cite{truccolo2005, pillow2008, london2010}, human activity \cite{sornette2004, barabasi2005}, and financial market activity \cite{fama1970, soros1987}.
Identifying whether the endogenous force, the exogenous force, or an interplay between the two is a major cause of the observed dynamics is important for the understanding of the system.
In the context of economics, a question of what causes movements in financial markets has been a central problem for a long time.
A classical paradigm of the efficient market hypothesis indicates that the market movements are fully governed by the arrival of news, that is, exclusively of exogenous origin \cite{fama1970}.
On the other hand, the empirical evidence has been accumulated that the news arrivals can explain only a part of the large movements, and that the market activity is thus largely of endogenous origin \cite{cutler1989, joulin2008}.
The endogenous effect of the markets is referred to as the reflexivity in the field of finance \cite{soros1987}.
To quantify the relative importance of each factor, appropriate modeling of the observed activity is necessary.

Filimonov and Sornette (2012) proposed to model the high-frequency financial data with a Hawkes process for quantifying the level of the reflexivity of the market  \cite{filimonov2012}.
A Hawkes process is a simple self-exciting point process to describe temporal clusters of events \cite{hawkes1971}.
In our situation, each event corresponds to a market movement.
In a Hawkes process, it is assumed that every event can trigger new events, and their occurrence rate is the sum of the background rate (exogenous effect) and the triggering effect from the preceding events (endogenous effect).
The strength of the endogeneity of the process is quantified by the average number of events that a single event directly triggers (the branching ratio).
By using a Hawkes model, it has been shown that a significant fraction of the market movements on the S$\&$P E-mini futures contracts is of endogenous origin \cite{filimonov2012, hardiman2013}.
In this way, the Hawkes process modeling is useful in analyzing the high-frequency financial data, and have become much popular in the domain of quantitative finance in this decade \cite{embrechts2011, bacry2013, sahalia2015} (also see references in \cite{bacry2015}).

In the original model of a Hawkes process \cite{hawkes1971}, the background rate is assumed to be constant in time, and many studies have employed this simple assumption.
However this assumption is not reasonable because the trading activity is non-stationary in general.
For example, it is well known that there is an intra-day seasonality, called U-shape pattern, where the activity is high around the opening and the closing of a session \cite{jain1988}.
In addition, an announcement of macroeconomic news, especially with surprising information, temporarily enhances the market activity \cite{petersen2010}, and it has been recently shown that the activity in a foreign exchange market after a macroeconomic news can be better described by a Hawkes model with a time-dependent background rate that exponentially decreases in time from the announcement \cite{rambaldi2015}.
Therefore it is necessary to consider the non-stationarity in the background rate to appropriately model the market dynamics.
There has been approaches to account for such a non-stationarity \cite{filimonov2012, hardiman2013, bowsher2007, lallouache2015, martins2016}.

In this paper, we develop a new Hawkes process model that can flexibly estimate the time-dependent background rate.
We demonstrate the effectiveness of our model by applying it to synthetic data and actual high-frequency data of Nikkei 225 mini futures contracts, and compare our model with existing models.
We also estimate the branching ratio of the market movements in Nikkei 225 mini and examine an impact of the non-stationarity on the estimate of the branching ratio.

\section{THE DATA}

\begin{figure}
 \includegraphics{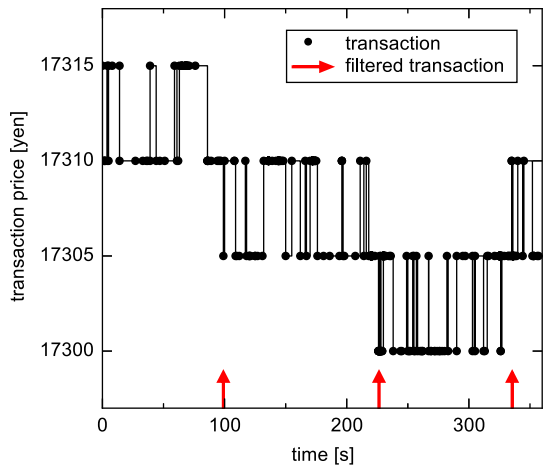}
 \caption{Example of the transaction price change (black dots) and the filtered transactions (red arrows; see text for the detail). The time series of the filtered transactions is analyzed in this study.}
 \label{fig-filtering}
\end{figure}

We use tick data of Nikkei 225 mini, a futures contract on the value of the Nikkei stock average, purchased from the JPX data cloud (http://db-ec.jpx.co.jp/).
Specifically, we analyze the data in the regular session (9:00-15:10 JST) from January 4, 2016 to June 30, 2016.
The dataset includes the time stamps, the volumes, and the prices of all the transactions.
The time stamps are recorded with a resolution of one second, and multiple events within the same interval of one second have the same time stamp.
To avoid the estimation bias from the rounding procedure, we add a uniform random number $[-0.5, 0.5(\mbox{seconds})]$ to each time stamp.
The tick size of the transaction price is five yen.
Nikkei 225 mini deals with the 16 contracts with different expiry dates.
For each session, we only use the most actively traded contract among them, the trade volume of which is $91\%$ of the total volume on average.

The change in the transaction price is not a good indicator of a market movement because the transaction price is subject to "microstructure" noise; the transaction price rapidly jumps back and forth between the best bid and ask prices in a liquid market (the bid-ask bounce; see Fig.~\ref{fig-filtering}) even if the best bid and ask prices remain the same.
Therefore, the change of the mid price, that is, the average of the best bid and ask prices, has been commonly used instead as a better indicator of the market movement \cite{filimonov2012, hardiman2013}.
However, our dataset does not include the information on the bid and ask prices.
Thus we alternatively filter the transactions for denoising the data in a following way so that the extracted transactions resemble the mid price changes in the occurrence pattern (Fig.~\ref{fig-filtering}).
We first exclude the transactions without the price change from the dataset.
We then select the transactions satisfying a condition that (i) the sign of the price change is the same as its previous transaction or (ii) the magnitude of the price change is greater than one tick size.
The filtered transactions account for about $3\%$ of all the transactions, and the average number per session is 2090, ranging from 465 to 18505.
In this study, we regard the filtered transaction as the market movement, and analyze the time series of the filtered transactions.

\section{MODEL}

\begin{figure*}
 \includegraphics{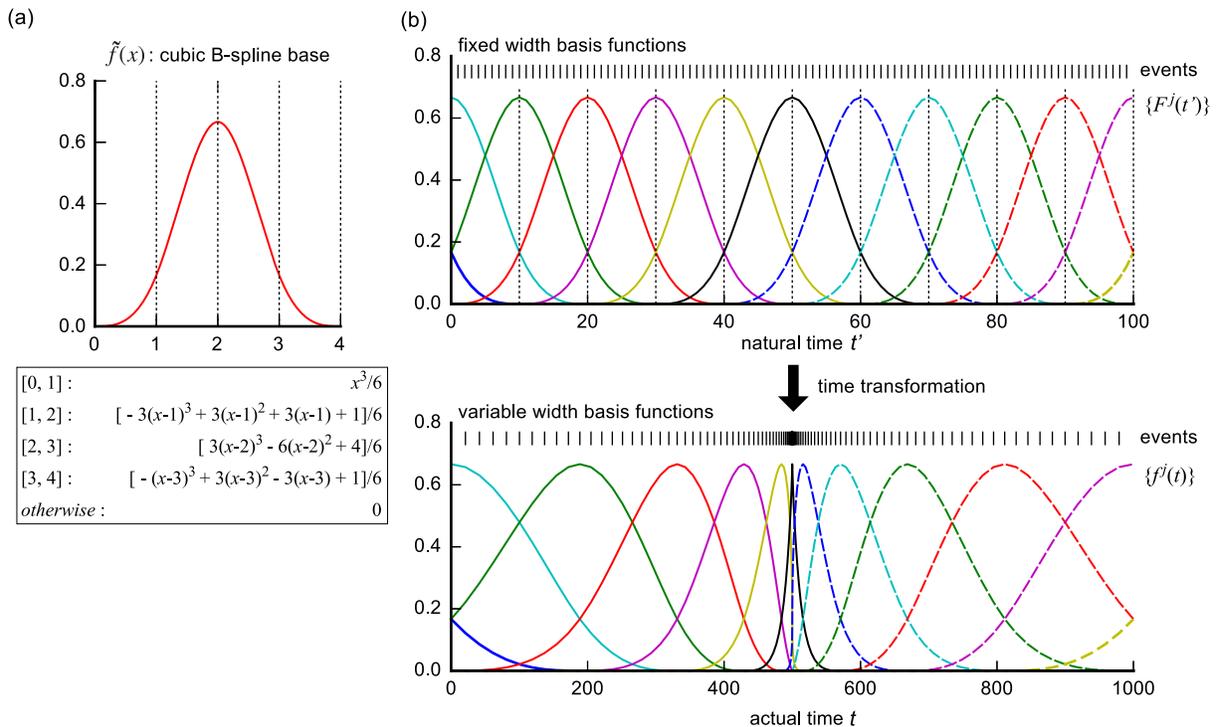}
 \caption{Schematic view of generating variable-width basis functions. See Appendix~\ref{App-A} for the detail method. (a) A cubic B-spline base. (b) We first place the events in an equidistant manner, and put the cubic B-spline bases with equal width represented by colored curves (the top panel). This time-coordinate is called natural time. The vertical dotted lines represent knots (See Appendixes~\ref{App-Z} and \ref{App-A}). We then make a coordinate change from the natural time to the actual time to obtain basis functions with variable width (the bottom panel). }
 \label{fig-cbs}
\end{figure*}

\begin{figure}
 \includegraphics{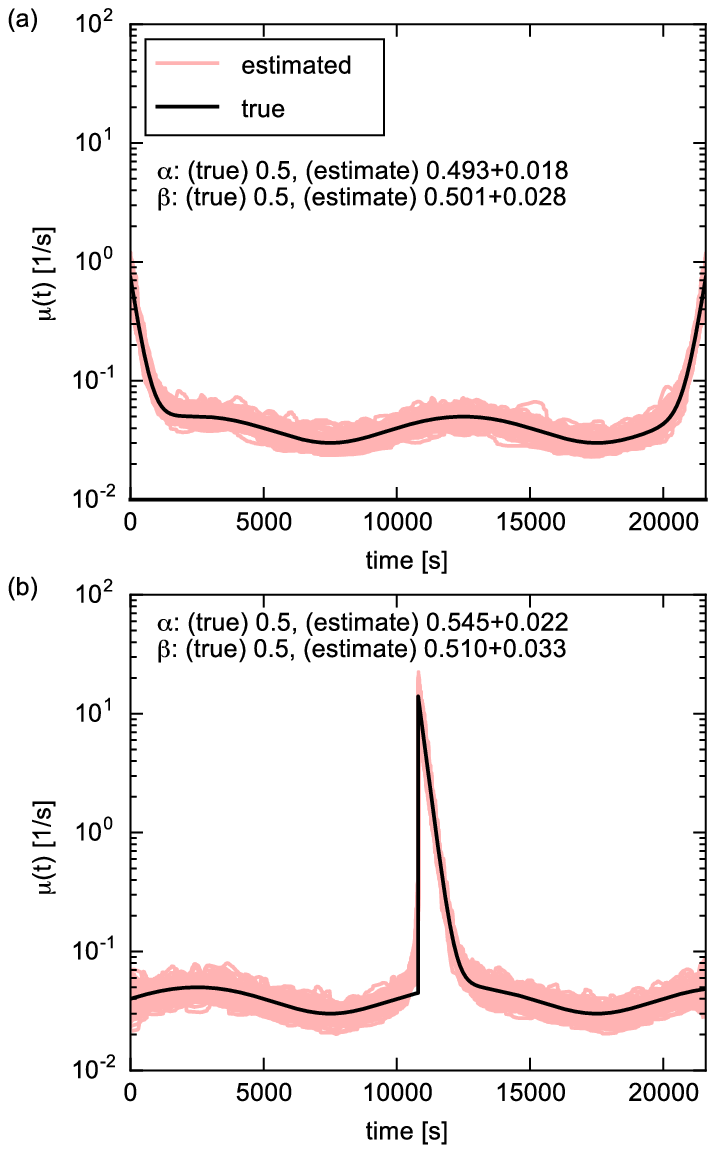}
 \caption{Simulation study. 100 synthetic sequences are simulated by using the Hawkes process with a given background rate $\mu(t)$ (black curve). Each red curve represents the estimate of the background rate obtained by applying our model to each sequence. Two different background rate functions are considered in (a) and (b).}
 \label{fig-sim}
\end{figure}

\subsection*{A. Hawkes process}

A Hawkes process is a simple point process model to describe clustering behavior of event occurrences \cite{hawkes1971}.
The original model and the extended models have been used in many fields \cite{ogata1988, mohler2011, masuda2013, reynaud2013}.
In a Hawkes process, the occurrence rate $\lambda(t)$ of events at time $t$ conditional on the occurrence history $H_t$ is given as
\begin{equation}
\lambda(t|H_t) = \mu(t) + \sum_{t_i<t}g(t-t_i),
\end{equation}
where the first term $\mu(t)$ is a background rate, and the second term represents the triggering effect from the preceding events at the occurrence times $t_i$, the strength of which is controlled by a triggering kernel function $g(\cdot)$. 
In a Hawkes process approach, we simply assume that the background/triggering term is exclusively attributed to the exogenous/endogenous effect.
The strength of the endogenity of the process is then characterized by the branching ratio $\int_{0}^{\infty}g(s)ds$, that is, the average number of events that are directly triggered by a single event.
The branching ratio is also interpreted as a ratio of the events with endogenous origin to all the events.

We use the sum of $M$ exponential functions for the triggering kernel function in this study as follows:
\begin{equation}
g(s) = \sum_{i=1}^{M}\alpha_i \beta_i e^{-\beta_i s},
\end{equation}
where $\alpha_i$ and $\beta_i$ $(i=1,\dots,M)$ are parameters.
In this parameterization, the sum of $\alpha$-parameters, $\sum_{i=1}^{M}\alpha_i$, corresponds to the branching ratio.
Although a power law function has been also used for the triggering kernel in the context of quantitative finance \cite{hardiman2013} to account for the long-range memory \cite{kwapien2012}, we here only consider a multiple exponential function because the previous study has shown that the model with a multiple exponential function is better fitted to the data than the one with a power law function \cite{lallouache2015}.

If we assume that the background rate is constant in time, $\mu(t)=\mu_c$, the model is characterized by the parameters $\theta=\{\mu_c,\boldsymbol{\alpha}, \boldsymbol{\beta}\}$, where $\boldsymbol{\alpha}=\{\alpha_1,\dots,\alpha_M\}$ and $\boldsymbol{\beta}=\{\beta_1,\dots,\beta_M\}$, and they can be estimated by a maximum likelihood method \cite{ozaki1979}.
The log-likelihood function of the Hawkes process of given parameters $\theta$ based on the data $\bold{D}^{[S,T]}=\{t_i\}_{i=1}^n$ of $n$ events in an observation interval $[S,T]$ is given as
\begin{equation}
\log L(\theta|\bold{D}^{[S,T]}) = \sum_{i=1}^{n}\log\lambda(t_i|H_{t_i}) - \int_S^T \lambda(t|H_t)dt.
\end{equation}

\subsection*{B. Our model with the non-stationary background rate}

We here develop a new Hawkes process model that can flexibly estimate the time-variation of the background rate $\mu(t)$. 
In this section, we only briefly summarize the method, and the details are described in Appendixes~\ref{App-Z}-\ref{App-C}.
In this study, $\log\mu(t)$ is modeled by a linear regression model with $m$ given basis functions $\{f^j(t)\}_{j=1}^m$ and parameters $\{a_j\}_{j=1}^m$ as follows:
\begin{equation}
\log \mu(t) = \sum_{j=1}^m a_j f^j(t). \label{eq:linearmodel}
\end{equation}
The linear model is considered for $\log \mu(t)$ since $\mu(t)$ only takes positive values.
In our model, the functional shapes and the number of the basis functions are determined based on the data for a flexible estimation from highly non-uniformly distributed data.
The basis functions $\{f^j(t)\}_{j=1}^m$ are bump shaped based on the cubic B-spline base $\tilde{f}(\cdot)$, which is a smoothly connected piecewise cubic function as shown in Fig.~\ref{fig-cbs}a.
In addition, the width of the basis functions is variable according to the frequency of the events to consider the non-uniformity of the event distribution: the width is small (large) for the high (low) frequency region (the bottom panel in Fig.~\ref{fig-cbs}b).
These variable-width basis functions are constructed by transforming the fixed width spline bases placed in the time coordinate (natural time) where the events are evenly spaced (Fig.~\ref{fig-cbs}b; see Appendixes~\ref{App-Z} and \ref{App-A} for the detail).
To make a flexible estimation, we prepare a relatively large number of the basis functions.
Specifically, the number of the basis functions $m$ is set to $3+\lfloor n/50 \rceil$, where $\lfloor \cdot \rceil$ represents the nearest integer function and $n$ is the number of events in a session.
How the performance depends on the number of the basis functions is summarized in Appendix~\ref{App-D}.

In our model, the number of the parameters of the basis functions $\{a_j\}_{j=1}^m$ is relatively large against the number of events (e.g. the order of hundreds for some data).
In such a case, a maximum likelihood estimation gives the rough estimates of $\mu(t)$ due to overfitting and tends to be computationally unstable.
To obtain a reliable estimation, we here employ a Bayesian framework with a smoothness constraint for $\mu(t)$ (see Appendix~\ref{App-B} for the detail of the estimation procedure). 
The Bayesian method has been commonly employed to estimate time-varying parameters from point process data \cite{ogata1993, paninski2010, omi2013}.
In the following, we refer to our new model as the Bayesian cubic B-spline (BCB) model. It is also noteworthy to mention that some models of a self-exciting point process with a non-stationary background rate have been proposed in other contexts \cite{mohler2011, kumazawa2014}.
The source code of our method is available upon request to the corresponding author.

To demonstrate the performance of our model, our model was applied to synthetic data simulated from the Hawkes process with a given time-varying background rate $\mu(t)$ (Fig.~\ref{fig-sim}).
In this test, we used a single exponential function for the triggering kernel ($M=1$).
The first example of $\mu(t)$ mimics the U-shape of the intraday pattern (Fig.~\ref{fig-sim}a).
We found that the estimates by our model well agreed with the true values of $\mu(t)$, $\alpha$, and $\beta$.
In the second example, we considered the discontinuous rise and following decay of the background activity, which mimics the market behavior after an announcement of important macroeconomic news (Fig.~\ref{fig-sim}b) \cite{rambaldi2015}.
Since $\mu(t)$ is modeled by a smooth function, our model cannot exactly reproduce such discontinuous behavior.
Actually, a close inspection of Fig.~\ref{fig-sim}b shows that the estimated $\mu(t)$ started to increase just before the true rise.
However, despite such a disadvantage, our method reasonably captured the trend of the true $\mu(t)$ owing to the use of the variable-width basis functions, and the bias in the estimate of $\alpha$ was slight.
These results showed the effectiveness of our model to flexibly estimate the time-varying background rate according to the data.

\subsection*{C. Existing studies of the non-stationary background rate}
We briefly review the existing studies which have dealt with the non-stationary background rate.
The simplest approach is to divide a session into sub-intervals of a short period (e.g. a few tens minutes) where the parameters can be considered to be constant and to separately apply the Hawkes process with a constant background rate to each subinterval \cite{filimonov2012}.
However if a rapid temporal variation of the background rate is present in a sub-interval, the estimation would be biased.
In addition, this approach requires each sub-interval to contain enough events for the model calibration, and the reliable estimation cannot be obtained for sub-intervals with not many events.

Bowsher (2007) has modeled the time-variation of the background rate by a piecewise linear function and assumed that the time-variation is identical for each business day \cite{bowsher2007}. 
Hardiman {\it et al.,} (2013) also have considered the detrending method that uses the identical weight function for each business day \cite{hardiman2013}.
However, as pointed out by Filimonov and Sornette (2015), the trading activity accompanies considerable day-to-day variability, then the assumption of the identical background rate across each session would result in the biased estimation \cite{filimonov2015}.

Some authors have modeled the background rate by using a piecewise linear function that is separately estimated for each session \cite{lallouache2015,martins2016}.
In this study, a piecewise linear function is used as the reference and compared with our model.
In the previous studies, knots are located at intervals of several hours \cite{lallouache2015,martins2016}, therefore we employ a knots location at two hours interval, [9:00, 11:00 ,13:00, 15:10].
We also consider a knots location at  30 minutes interval, [9:00, 9:30, 10:00,..., 14:30, 15:10].
The parameters of this model are estimated by a maximum likelihood method.

\section{RESULTS}

\subsection*{A. Comparison between the Hawkes models with different background rate functions}

\begin{table}
\caption{The average fitting score per session. Each value is relative to the best model, the BCB model with $M=2$.}
\label{table-score}
\begin{ruledtabular}
\begin{tabular}{c|cccc}
& \multicolumn{4}{c}{The number of exponentials ($M$)} \\
& 1 & 2 & 3 & 4 \\
\hline
CONST & -110.8 & -28.3 & -17.9 & -18.3 \\
$\mbox{PL}_{\rm 2h}$ & -77.2 & -18.5 & -13.7 & -14.6 \\
$\mbox{PL}_{\rm 30min}$ & -47.4 & -7.7 & -7.8 & -9.5 \\
BCB & -19.7 & 0.0 & -0.3 & -1.9 \\
\end{tabular}
\end{ruledtabular}
\end{table}

\begin{figure}
 \includegraphics{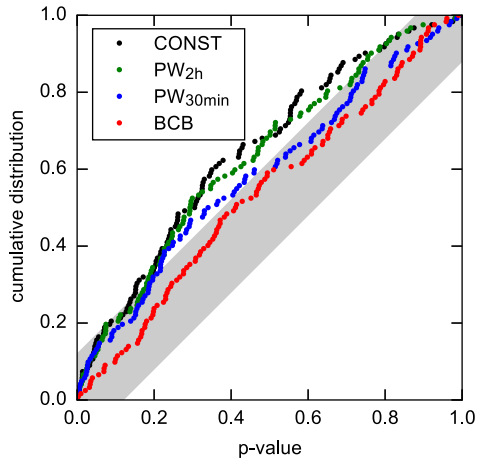}
 \caption{The cumulative distribution of the $p$-value obtained in the KS test for each session for the CONST, $\mbox{PL}_{\rm 2h}$ $(M=3)$, $\mbox{PL}_{\rm 30min}$ $(M=2)$, and BCB $(M=2)$ models. If a distribution is in a gray zone, the distribution passes the KS test of the uniformity with $p=95\%$.}
 \label{fig-ks}
\end{figure}

\begin{figure*}
 \includegraphics{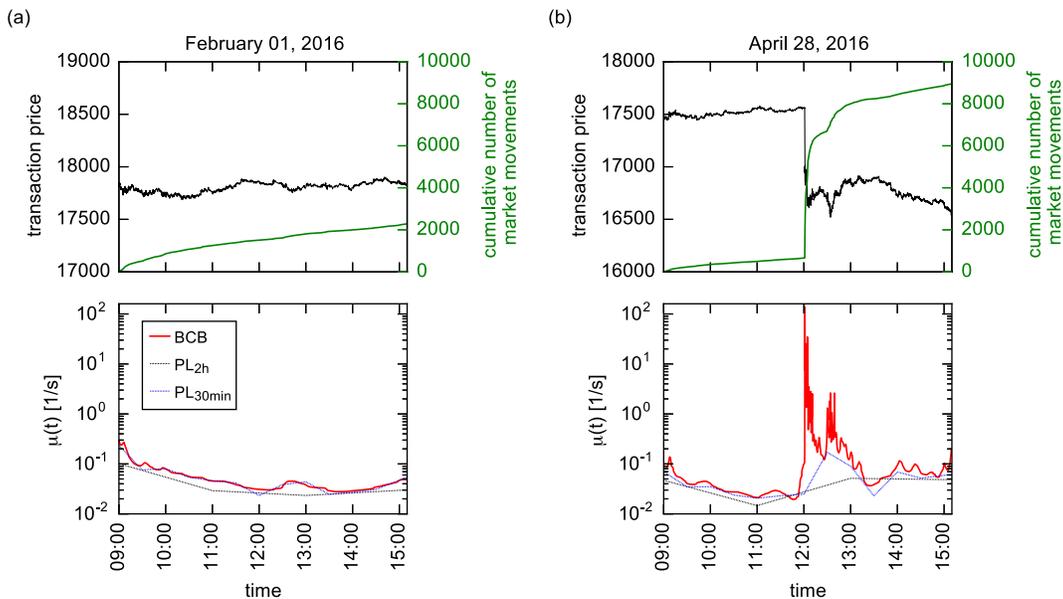}
 \caption{Examples of the estimation of the background rate from the data of the Nikkei 225 mini for the sessions on (a) February 01, 2016 and (b) April 28, 2016. The estimates from the BCB ($M=2$), $\mbox{PL}_{\rm 2h}$ ($M=3$), and $\mbox{PL}_{\rm 30min}$ ($M=2$) models are shown, where $M$ represents the number of exponentials in the triggering kernel. The fitting scores ($\mbox{PL}_{\rm 2h}$/$\mbox{PL}_{\rm 30min}$/BCB) are respectively -5.2/0.0/0.0 for (a) and -303.1/-299.5/0.0 for (b), where each value is relative to the BCB model. The branching ratios are respectively 0.62/0.49/0.46 for (a) and 0.91/0.87/0.48 for (b).}
 \label{fig-ex-mu}
\end{figure*}

\begin{figure*}
 \includegraphics{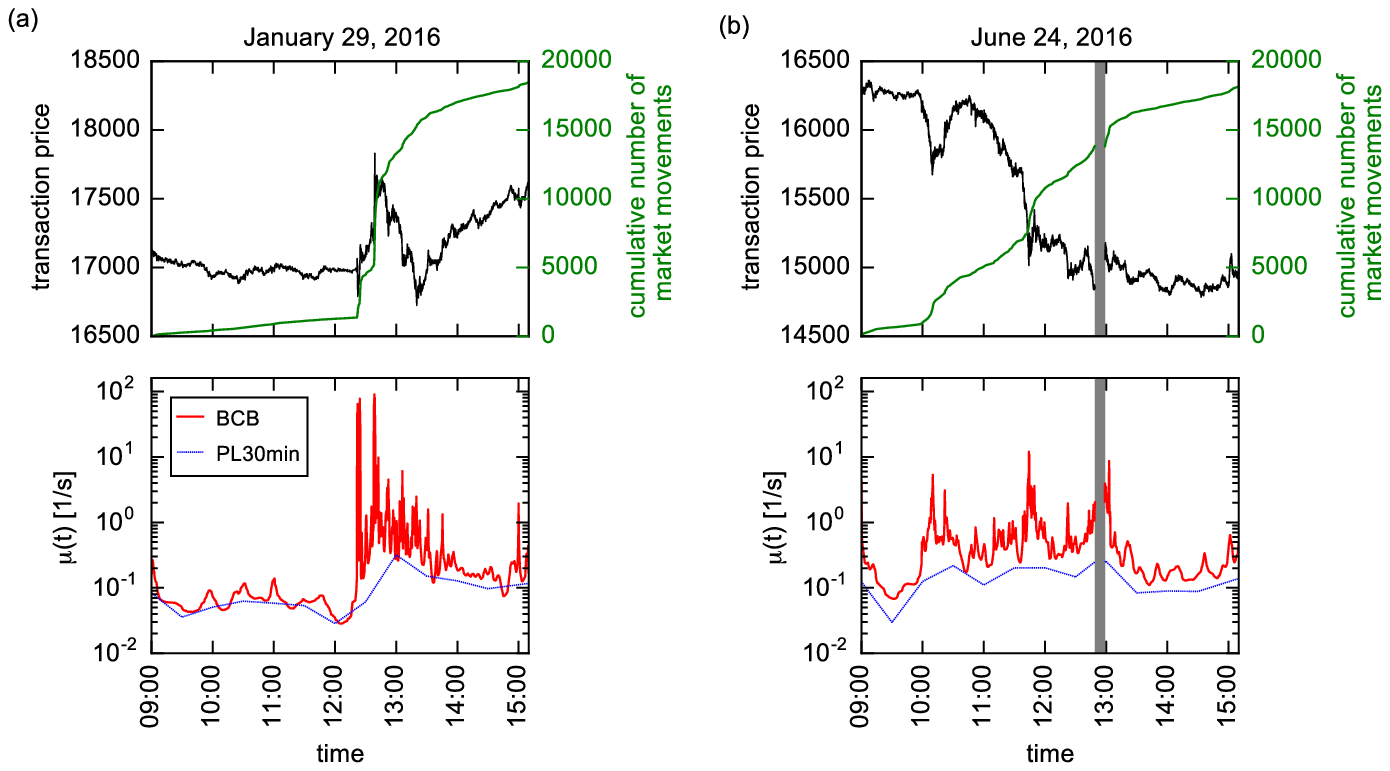}
 \caption{Examples of the estimation of the background rate from the data of the Nikkei 225 mini for the sessions on (a) January 29, 2016, when there was an announcement by Bank of Japan, and (b) June 24, 2016, when the United Kingdom European Union membership referendum resulted in the UK voting to leave the EU. The estimates from the $\mbox{PL}_{\rm 30min}$ ($M=2$) and BCB ($M=2$) models are shown. The fitting scores ($\mbox{PL}_{\rm 30min}$/BCB) are respectively -342.2/0.0 for (a) and -77.1/0.0 for (b), where each value is relative to the BCB model. The branching ratios are respectively 0.89/0.54 for (a) and 0.84/0.47 for (b). In a gray shaded zone in (b), the trading was stopped due to the circuit breaker.}
 \label{fig-ex-mu2}
\end{figure*}

We compared the performance of the Hawkes models with four different background rate functions: (i: CONST) a constant background rate, (ii: $\mbox{PL}_{\rm 2h}$) the piecewise linear model with a knot interval of two hours, (iii: $\mbox{PL}_{\rm 30min}$) the piecewise linear model with a knot interval of 30 minutes, and (iv: BCB) our new model of the Bayesian cubic B-spline model.
We fitted each model to the data in the regular session of 6 hours and 10 minutes on each business day, and obtained the separate estimates for the total of 122 business days.

To quantify the goodness-of-fit to the data of the models with the different levels of complexity, we used the Akaike information criterion (AIC) \cite{akaike1974} for the CONST, $\mbox{PL}_{\rm 2h}$, and $\mbox{PL}_{\rm 30min}$ models and the Akaike Bayesian information criterion (ABIC) \cite{akaike1980} for the BCB model (see Appendix~\ref{App-C} for the detail).
For each session, the score of each model, $-AIC/2$ or $-ABIC/2$ (the higher value means the better fit), was calculated.
The score takes the form $S = \log (Likelihood) - (Penalty)$, where the penalty depends on the model complexity (see Appendix~\ref{App-C} for the detail).
Table~\ref{table-score} summarizes the average score per session of each model.
We found that the best model was the BCB model with the two exponentials, and that the BCB model with the three exponentials also performed similarly to the best model, demonstrating the improvement in the goodness-of-fit by our model over the other models.
The fitting was generally improved as flexibility of the background rate function is increased (from top to bottom in the Table~\ref{table-score}).
We also found that the model with multiple exponentials was significantly better than the one with a single exponential, which is consistent with the previous study \cite{lallouache2015}.
In the following analysis, we only focus on the CONST ($M=3$), $\mbox{PL}_{\rm 2h}$ ($M=3$), $\mbox{PL}_{\rm 30min}$ ($M=2$), and BCB ($M=2$) models, which are respectively the best for each class of the background rate function.

Next, we checked whether the estimated models were statistically consistent with the data.
If a model is correct, the transformed inter-event intervals $\{\tau_i = 1-\exp{[-\int_{t_i}^{t_{i+1}}\lambda(t)dt]}\}$ follow the uniform distribution.
The Kolmogorv-Smirnov (KS) test under the null hypothesis of the uniform distribution was applied to $\{\tau_i\}$ for each session to obtain the $p$-value.
Figure~\ref{fig-ks} shows the cumulative distribution of the $p$-value over all the sessions.
The distribution of the $p$-value also follows the uniform distribution under the null hypothesis, then we again applied the KS test to the set of the obtained $p$-values to check its uniformity.
The BCB model passed the test with the confidence level of $95\%$, but the CONST, $\mbox{PL}_{\rm 2h}$, and $\mbox{PL}_{\rm 30min}$ models were rejected (Fig.~\ref{fig-ks}).
This result showed the consistency of the BCB model with the data.

Figure~\ref{fig-ex-mu} shows the examples of the estimates of the background rate $\mu(t)$ associated with dynamics of the price and the cumulative number of the market movements (the filtered transactions) for the two contrasting sessions.
While the price was almost flat in the first example (Fig.~\ref{fig-ex-mu}a), a sharp decline in the price was present in the second example because Bank of Japan announced to keep monetary policy unchanged and disappointed the market (Fig.~\ref{fig-ex-mu}b).
We here compared the performances between the $\mbox{PL}_{\rm 2h}$, $\mbox{PL}_{\rm 30min}$, and BCB models.
In both cases, the $\mbox{PL}_{\rm 2h}$ model performed worse than the other models.
In the first example, the $\mbox{PL}_{\rm 30min}$ and BCB models gave the similar estimates, where the background rate fluctuated slowly and slightly.
Therefore the two models could perform similarly for sessions without large price change.
To examine this result, we chose the sessions where the range of the price, the difference between the maximum and minimum price during a session, is equal to or less than its $75\%$ quantile (385 yen).
For such sessions without large price change, the average score of the $\mbox{PL}_{\rm 30min}$ model was slightly higher than the BCB model by $0.3$.
In addition, the $\mbox{PL}_{\rm 30min}$ model passed the KS test for this case.
Thus, in addition to the BCB model, the $\mbox{PL}_{\rm 30min}$ is also effective to analyze sessions where the background rate slowly varies in time.
 
In contrast, in the second example, the $\mbox{PL}_{\rm 30min}$ and BCB models performed quite differently and the BCB model was significantly better than the $\mbox{PL}_{\rm 30min}$ model.
For this particular example, the difference in the score was about 300, implying the BCB model was $e^{300}$ times more probable than the $\mbox{PL}_{\rm 30min}$ model.
The BCB model showed the sudden rise and decay of the background rate along with the news announcement, which is consistent with the results in Rambaldi et al (2015) \cite{rambaldi2015}, while the $\mbox{PL}_{\rm 30min}$ model cannot capture such fine time-variation.
These results suggest that it is important for a model to take into account the fine-scale temporal variation of the background rate and that the BCB model is particularly useful in analyzing sessions where the background rate fluctuates considerably.
It should be also noted that the form of the decay was not so simple: the background rate increased again about 30 minutes after the announcement.
Other examples where the considerable time-variation of the background rate was present are shown in Fig.~\ref{fig-ex-mu2}.
The market response to a macroeconomic news announcement is quite variable in a sense that even the same news can cause the different response.
At this point, the BCB model may be also useful to characterize the complex market response to a given news announcement.

\subsection*{B. Impact of the non-stationarity of the background rate on the estimate of the branching ratio}

\begin{table}
\caption{The average branching ratio.}
\label{table-br}
\begin{ruledtabular}
\begin{tabular}{c|cccc}
& \multicolumn{4}{c}{The number of exponentials ($M$)} \\
& 1 & 2 & 3 & 4 \\
\hline
CONST & 0.57 & 0.74 & 0.81 & 0.83 \\
$\mbox{PL}_{\rm 2h}$ & 0.43 & 0.59 & 0.66 & 0.68 \\
$\mbox{PL}_{\rm 30min}$ & 0.32 & 0.45 & 0.47 & 0.49 \\
BCB & 0.25 & 0.41 & 0.47 & 0.49 \\
\end{tabular}
\end{ruledtabular}
\end{table}

Table~\ref{table-br} summarizes the average branching ratio per session for each model.
The average branching ratio estimated from the best model (the BCB model with two exponentials) was 0.41, implying that $41\%$ of the market movements was of endogenous origin in Nikkei 225 mini.
This result shows the relative importance of the exogenous factor in the market dynamics, which is in contrast to the previous studies that have argued the significant contribution of the endogenous effect \cite{filimonov2012, hardiman2013}.

We examined the impact of the non-stationarity of the background rate on the estimate of the branching ratio.
The branching ratio was clearly overestimated for the CONST and $\mbox{PL}_{\rm 2h}$ models as compared to the best model (the BCB model).
The CONST and $\mbox{PL}_{\rm 2h}$ models cannot capture the time-variation of the background rate well due to its low flexibility as shown in Fig.~\ref{fig-ex-mu}, and then try to reproduce the non-stationarity in the data as a consequence of the self-excitation, resulting in the overestimation of the branching ratio.
The $\mbox{PL}_{\rm 30min}$ ($M=2$) model gave the branching ratio, on average, similar to the one by the BCB ($M=2$) model.
Thus we have to consider the temporal variation of the background rate at least at the time scale of a few tens minutes to appropriately estimate the branching ratio.
On the other hand, The $\mbox{PL}_{\rm 30min}$ ($M=2$) model overestimated the branching ratio for sessions where the considerable variation of the background rate was present (Figs.~\ref{fig-ex-mu}b and \ref{fig-ex-mu2}) due to its inability to reproduce the rapid temporal variation.
In contrast, the BCB model can estimate the branching ratio more accurately by flexibly estimating the background rate.
These results demonstrates that the appropriate modeling of the background rate is critically important for the branching ratio estimation and that the use of the overly smooth background rate results in the overestimation of the branching ratio.

\subsection*{C. Day-to-day variability of the background rate}

\begin{figure}
 \includegraphics{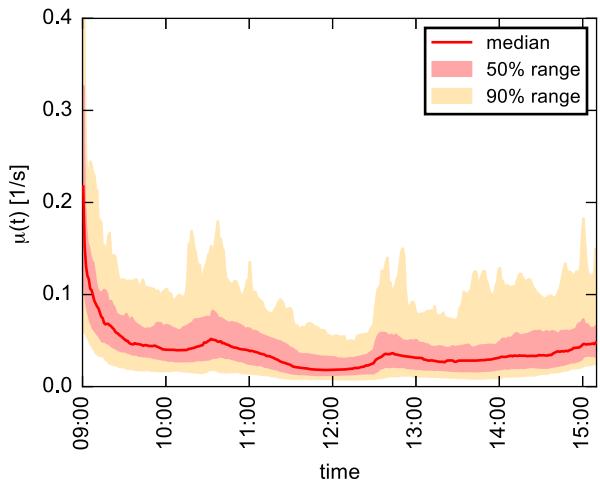}
 \caption{The day-to-day distribution of the background rate from January 04, 2016 to June 30, 2016,  estimated by the BCB ($M=2$) model.}
 \label{fig-dist-mu}
\end{figure}

To examine the intraday pattern in the background rate, the median, $50\%$, and $90\%$ ranges of the background rate at each time over the all sessions under consideration are plotted (Fig.~\ref{fig-dist-mu}).
The background rate was clearly high around the opening of a session, especially for the first 30 minutes, but there was no clear increase around the closing of a session.
This tendency was also observed for the occurrence rate of the market movements, and that of all the transactions including the ones without price change. 
Thus the market movement in Nikkei 225 mini exhibited the reverse J-shaped intraday pattern rather than the U-shaped pattern.
We also found that there was considerable day-to-day variability in the background rate.
Therefore, it is necessary to model the background rate in an adaptive way.

\section{CONCLUSION AND DISCUSSION}

In this paper, we have developed the Hawkes model with a time-dependent background rate, and applied it to the high-frequency financial data of Nikkei 225 mini.
Our model can flexibly estimate not only the slow variation but also the rapid variation in the background rate owing to the use of a relatively large number of the variable-width basis functions.
Our model was better fitted to the data than the Hawkes models with other background rate functions, the constant model (CONST), the piecewise linear model with a knot interval of 2 hours ($\mbox{PL}_{\rm 2h}$), and the piecewise linear model with a knot interval of 30 minutes ($\mbox{PL}_{\rm 30min}$).
Especially the improvement in the goodness-of-fit to the data by our model was prominent for sessions where the considerable temporal variation of the background rate was present.
Furthermore, we have found that our model was statistically consistent with the data.

We have demonstrated the importance of appropriately modeling the time-dependent background rate for the branching ratio estimation.
The CONST and $\mbox{PL}_{\rm 2h}$ models generally overestimated the branching ratio because they cannot capture the time-variation of the background rate well.
The branching ratio estimated by the $\mbox{PL}_{\rm 30min}$ model was, on average, similar to that by our model, implying that it is necessary to consider the temporal variation at least at this time-scale.
However, the $\mbox{PL}_{\rm 30min}$ model overestimated the branching ratio if the rapid fluctuation of the background rate was present in a session.
In contrast, our model can estimate the branching ratio more accurately by flexibly estimating the time-variation of the background rate.
We also have revealed that the average branching ratio for the Nikkei 225 mini estimated by our model is 0.41, suggesting the relative importance of the exogenous factor in contrast to the previous studies \cite{filimonov2012,hardiman2013}.

Generally it is a difficult task to decompose the origin of the event occurrences into the endogenous and exogenous effects given the unknown time-varying background rate \cite{filimonov2015}.
The bias in the estimate of the background rate would lead to the bias in the estimate of the branching ratio.
For example, the use of an overly smooth background rate leads to the overestimation of the branching ratio, as shown above.
To reduce subjectivity in this problem, we have objectively made this decomposition according the data so that the model's ability to reproduce the data is maximized.
Specifically, the parameters of the triggering kernel, the degree of the smoothness of the temporal variation, and the time-course of the background rate were simultaneously optimized by maximizing the marginal likelihood function to the data.
In addition, we have confirmed that the fitting was improved by our model over the other models and that our model was statistically consistent with the data.

Our method to estimate the time-dependent background rate is not suitable for the online estimation.
We employed the smoothing method: the value at a given time is estimated using the data around the time.
On the other hand, only the data before the given time have to be used for the online estimation.
For such a purpose, a non-gaussian state-space model would be useful \cite{kitagawa1987}.


\section*{ACKNOWLEDGEMENTS}
This work is supported by the Kozo Keikaku Engineering Inc. and CREST, JST.
We would like to thank two anonymous referees for their useful comments for the significant revision of the present paper.

\appendix

\section{B-spline curve estimation approach}
\label{App-Z}
Our model of a time-dependent background rate is essentially based on the B-spline curve estimation \cite{deboor1978}.
Let us consider a target curve $h(t)$ in $[0,T]$ and equally spaced knots at $t=\eta_l \equiv Tl/m'$ for a given integer $m'$ and integers $l$, which divide the observation interval $[0,T]$ into $m'$ sub-intervals $\{[\eta_{l-1},\eta_{l}]\}_{l=1}^{m'}$ of equal length of $w'=T/m'$.
In this approach, the function $h(t)$ is modeled using a function that is a cubic polynomial in each sub-interval and is smooth at knots.
Such a function is efficiently obtained by a linear combination of $(m'+3)$ fixed-width basis functions $\{q^j(t)=\tilde{f}((t-\eta_{j-4})/w')\}_{j=1}^{m'+3}$, where $\tilde{f}(\cdot)$ is a cubic B-spline base in Fig.~\ref{fig-cbs}a.
Here the basis function $q^j(t)$ has a support on $[\eta_{j-4},\eta_{j}]$, and is a smooth piecewise cubic polynomial of bump shape.

Although a liner combination of the fixed-width B-spline basis functions has a certain degree of flexibility in representing a variety of curves, this method is not suitable for describing the time-variation of the background rate where the rapid temporal variation in a short period is present.
A basis function cannot capture the rapid temporal variation accompanied with the burst of events unless we employ an extremely large number of the basis functions.
On the other hand, since the events in high-frequency financial data are distributed in a highly non-uniform way, a support of a basis function for a low activity region contains only a small number of the events leading to a poor estimate of the corresponding parameter.
To fix this problem, we use variable width basis functions where the width depends on the frequency of the events as in the bottom panel of Fig.~\ref{fig-cbs}b.

\section{Constructing variable-width basis functions}
\label{App-A}

A specific method to generate bump-shaped basis functions with variable width is described.
Our method is schematically illustrated in Fig.~\ref{fig-cbs}b.
We first introduce a natural time coordinate $t'$, where the position $t'_i$ of the $i$th event is equal to the index, $t'_i=i\ (i=1,2,\dots,n)$, and the start $t'_0$ and end $t'_{n+1}$ of the observation interval are respectively $0$ and $n+1$ (the top panel in Fig.~\ref{fig-cbs}b). 
We then obtain $m$ fixed-width basis functions $\{F^j(t')=\tilde{f}((t'-\xi_{j-4})/w)\}_{j=1}^{m}$ as in the Appendix~\ref{App-Z} (the top panel in Fig.~\ref{fig-cbs}b), where we place knots at ${t = \xi_l \equiv (n+1)l/(m-3)}$ in an interval of $w=(n+1)/(m-3)$.

Next, we make a coordinate change from the natural time $t'$ to the actual time $t$ to obtain the basis functions $\{f^j(t)\}_{j=1}^{m}$ with variable width (Fig.~\ref{fig-cbs}b).
Although the basis functions $\{f^j(t)\}_{j=1}^{m}$ are continuous functions under appropriate time-transformation, we simply approximate the basis functions $\{f^j(t)\}_{j=1}^{m}$ to be constant during each inter-event period.
Accordingly the basis functions in the actual time coordinate are given as $f^j(t)=f_i^j \equiv F^j(t'_i)$ for $t_i \leq t < t_{i+1} \ (i=0,1,\dots,n)$, where $t_0 = S$ and $t_{n+1} = T$.
This approximation also makes the calculation of the log-likelihood function easy.

\section{Bayesian estimation of the Hawkes process model with the time-dependent background rate}
\label{App-B}

Our model is characterized by the parameters $\{\bold{a}, \boldsymbol{\alpha}, \boldsymbol{\beta}\}$, where $\bold{a}=(a_1,\dots,a_m)^T$, and the log likelihood function is given as 
\begin{widetext}
\begin{equation}
\log L(\bold{a}, \boldsymbol{\alpha}, \boldsymbol{\beta}|\bold{D}^{[S,T]}) = \sum_{i=1}^{n}\log \left[ \mu_i + \sum_{k=1}^{i-1} \sum_{j=1}^{M}\alpha_j \beta_j e^{-\beta_j(t_i-t_k)} \right] - \sum_{i=0}^n \mu_i(t_{i+1}-t_i)- \sum_{i=1}^n \sum_{j=1}^{M} \alpha_j [1-e^{-\beta_j(T-t_i)}],
\end{equation}
\end{widetext}
where $\mu_i = \exp[\sum_{j=1}^{m}a_i f_i^j]$.
In our case, the number of the parameters of $\bold{a}$ is relatively large, the order of hundreds for some data analyzed in this study.
In such a case, the maximum likelihood estimation gives a rough estimate of $\mu(t)$ due to overfitting and tends to be computationally unstable.
To robustly estimate such a large number of the parameters, we employ a Bayesian estimation approach.
In the approach, we introduce a prior probability distribution of the parameters $\bold{a}$ that restricts the flexibility of $\mu(t)$ to prevent overfitting, given as
\begin{widetext}
\begin{eqnarray}
P_{V, W, \mu_c}(\bold{a}) = \frac{1}{C} \exp\left[-\frac{V}{2}\sum_{i=1}^n\left(\frac{d}{dt'}\sum_{j=1}^{m}a_jF^j(t'_i)\right)^2 - \frac{W}{2}\left(\frac{\sum_{j=1}^{m}a_j}{m}-\log\mu_c \right)^2 \right], 
\label{eq_prior}
\end{eqnarray}
\end{widetext}
where $C$ is the normalizing constant, and $V$, $W$, and $\mu_c$ are additional parameters.
The first term in the exponential penalizes the first derivative of $\log\mu(t')=\sum_{j=1}^{m}a_j F^j(t')$ in the natural time coordinate $t'$, and the parameter $V$ controls the smoothness of $\mu(t)$ (also see Appendix~\ref{App-C}).
The second term in the exponential restricts the average parameter value $\sum_{j=1}^{m}a_j/m$, which roughly corresponds to the baseline of $\log\mu(t)$, $\sum_{i=1}^{n}\log\mu(t_i)/n$.
This term is introduced to make the prior normalizable; since the first term is invariant under the transformation $a_j \to a_j + c\ (j=1,\dots,m)$ for constant $c$, which corresponds to the vertical shift $\log\mu(t) \to \log\mu(t)+c$ from the relation $\sum_{j=1}^m f_i^j=1$, the prior only with the first term cannot be normalized.
In this study, the parameter $W$ is fixed to a very large value (e.g. $10^4$) because the estimate of the model parameters $\{\bold{a}, \boldsymbol{\alpha}, \boldsymbol{\beta}\}$ remain nearly unchanged regardless of the value of $W$.
In this setting, we have a relation $\log\mu_c \simeq \sum_{i=1}^{n}\log\mu(t_i)/n$, so the parameter $\mu_c$ represents the baseline of $\mu(t)$.
We also note that our non-stationary model reduces to the stationary process with $\mu(t) = \mu_c$ in the limit of $V \to \infty$ and $W \to \infty$.

In a Bayesian framework, we treat the parameters $\theta_h = \{\boldsymbol{\alpha}, \boldsymbol{\beta}, V, \mu_c\}$ as hyper-parameters, and estimate the hyper-parameters $\theta_h$ differently from the remaining parameters $\bold{a}$.
We first consider a method to estimate the parameters $\bold{a}$ under a given value of the hyper-parameters $\theta_h$.
From the Bayes' theorem \cite{bishop2006}, the posterior probability distribution of the parameters $\bold{a}$ given the hyper-parameters $\theta_h$ is given by
\begin{equation}
P_{\theta_h}(\bold{a}|D^{[S,T]}) \propto L(\bold{a},\boldsymbol{\alpha}, \boldsymbol{\beta}|\bold{D}^{[S,T]}) P_{V, \mu_c}(\bold{a}).
\end{equation}
The maximum a posteriori (MAP) estimate $\bold{a}^*$ is obtained by maximizing the logarithm of the posterior.

Then we consider to find the optimal estimate of the hyper-parameters $\theta_h$.
The goodness-of-fit to the data of the model with the hyper-parameters $\theta_h$ is characterized by the marginal likelihood function, also known as the evidence, given as
\begin{equation}
ML({\theta_h}|D^{[S,T]}) = \int L(\bold{a}, \boldsymbol{\alpha}, \boldsymbol{\beta}|\bold{D}^{[S,T]}) P_{V, \mu_c}(\bold{a}) d\bold{a}.
\end{equation}
We approximately calculate the marginal likelihood function by using Laplace's method \cite{tierney1986}; the logarithm of the integrand is expanded to the second order of $\bold{a}$ around its peak.
In our case, the peak is located at $\bold{a}=\bold{a}^*$, which is the MAP estimate.
We then obtain
\begin{widetext}
\begin{equation}
\log ML({\theta_h}|D^{[S,T]}) \simeq \frac{m}{2}\log 2\pi - \frac{1}{2}\log\det(H) + \log L(\bold{a^*}, \boldsymbol{\alpha}, \boldsymbol{\beta}|\bold{D}^{[S,T]}) + \log P_{V, \mu_c}(\bold{a^*}), \label{eq-ml}
\end{equation}
\end{widetext}
where $H$ represents the Hessian matrix of the minus of the logarithm of the integrand at $\bold{a}=\bold{a^*}$.
The estimate $\theta_h^*$ of the hyper-parameters is obtained by maximizing the logarithm of the marginal likelihood function \cite{good1965}.
The optimal estimate of the parameter set $\{\bold{a}^*,\theta_h^*\}$ is finally obtained as the optimal estimate of the hyper parameters $\theta_h^*$ and the MAP estimate $\bold{a}^*$ under the optimal hyper-parameters $\theta_h^*$.

\section{Model Selection}
\label{App-C}

In the text, we compare the goodness-of-fit to the data of the Hawkes process models with different background rate models.
The log-likelihood is not a good measure of the goodness-of-fit in comparing models with different levels of complexity, because the log-likelihood of a complex model overestimates the goodness-of-fit due to overfitting.
Therefore it is a standard task to add a penalty on the log likelihood depending on the model complexity \cite{akaike1974, schwarz1978}.
For a model that is estimated in a maximum likelihood method (the CONST, $\mbox{PL}_{\rm 2h}$, and $\mbox{PL}_{\rm 30min}$ models), we use a score $\arg \max_{\theta} \log L(\theta|D) - \mbox{\it (the number of the parameters)}$ based on the Akaike information criterion \cite{akaike1974}, which corresponds to $-AIC/2$.
For our model that is estimated in the Bayesiam way, we use a score $\arg \max_{\theta_h} \log ML(\theta_h|D) - \mbox{\it (the number of the hyper-parameters)}$ based on the Akaike Bayesian information criterion \cite{akaike1980}, where $ML(\theta_h|D)$ is the marginal likelihood in Eq.~(\ref{eq-ml}).
This score corresponds to $-ABIC/2$.
It is noted that the penalty on the model complexity coming from the parameters $\bold{a}$ that account for the time variation of $\mu(t)$ is naturally included in the log marginal likelihood.
We can rewrite Eq.~(\ref{eq-ml}) as $\log ML = \log L - (Penalty)$.
The complexity of the temporal variation in $\mu(t)$ is controlled by the hyper-parameter $V$; the estimated $\mu(t)$ is almost constant for a large value of $V$, and the estimated $\mu(t)$ much varies in time for a small value of $V$.
The penalty term is almost 0 for the large value of $V$ (low complexity), and scales with $m \log (1/V)$ for the small value of $V$ (high complexity).

\section{The dependence of the performance on the number of the basis functions.}
\label{App-D}

\begin{table}[b]
\caption{The average fitting score ($-ABIC/2$) of our model per session as a function of the number of basis functions $m=3+\lfloor n/k \rceil$ and the number of exponential functions $M$. Each value is relative to the best model ($k=50$, $M=2$).}
\label{table-nb}
\begin{ruledtabular}
\begin{tabular}{c|cccc}
& M=1 & M=2 & M=3 & M=4 \\
\hline
k=50 & -19.7 & 0.0 & -0.3 & -1.9 \\
k=75 & -26.4 & -2.6 & -2.1 & -3.2 \\
k=100 & -31.7 & -5.1 & -4.2 & -5.6 \\
k=150 & -38.7 & -7.9 & -6.3 & -7.3 \\
k=200 & -43.9 & -10.2 & -7.9 & -9.0 \\
\end{tabular}
\end{ruledtabular}
\end{table}

Table~\ref{table-nb} summarizes the dependence of the goodness-of-fit of our model on the number of basis functions. The fitting was generally improved by increasing the number of basis functions for smaller $k$.
We also checked the performance with $k=25$, but the estimation was computationally unstable for some data due to a quite large number of the parameters.

\end{document}